
\input phyzzx
\vsize=23.5 true cm
\hsize=15.4 true cm
\predisplaypenalty=0
\abovedisplayskip=3mm plus6pt minus 4pt
\belowdisplayskip=3mm plus6pt minus 4pt
\abovedisplayshortskip=0mm plus6pt
\belowdisplayshortskip=2mm plus6pt minus 4pt
\normalbaselineskip=12pt
\normalbaselines
\noindent
\def\e{\, {\rm e}}
\vsize=23.5 true cm
\hsize=15.4 true cm

\REF\BKDSGM{E. Br\'ezin and V.A. Kazakov,
{\it Phys.\ Lett.\ }{\bf 236B} (1990) 144;
M. Douglas and A. Shenker, {\it Nucl.\ Phys.\ }{\bf B335} (1990)
635; D.J. Gross and A.A. Migdal, {\it Phys. Rev. Lett.\ }{\bf 64}
(1990) 127.}
\REF\GM{D.J. Gross and A.A. Migdal,
{\it Nucl.\ Phys.\ }{\bf B340} (1990) 333;
T. Banks, M. Douglas, N. Seiberg and S. Shenker,
{\it Phys.\ Lett.\ }{\bf 238B} (1990) 279.}
\REF\KPZ{V.G. Knizhnik, A.M. Polyakov and A.B. Zamolodchikov,
{\it Mod.\ Phys.\ Lett.\ }{\bf A3} (1988) 819.}
\REF\DDK{F. David, {\it Mod.\ Phys.\ Lett.\ }{\bf A3} (1988) 1651;
J. Distler and H. Kawai, {\it Nucl.\ Phys.\ }{\bf B321} (1989) 509.}
\REF\SEPO{N. Seiberg,
{\it Prog.\ Theor.\ Phys.\ Suppl.\ }{\bf 102} (1990) 319;
J. Polchinski,
in {\it Strings '90}, eds. R. Arnowitt {\it et al}.
(World Scientific, Singapore, 1991).}
\REF\KITA{M. Goulian and M. Li,
{\it Phys.\ Rev.\ Lett.\ }{\bf 66} (1991) 2051;
Y. Kitazawa,
{\it Phys.\ Lett.\ }{\bf 265B} (1991) 262;
{\it Int.\ J.\ Mod.\ Phys.\ }{\bf A7} (1992) 3403;
K. Aoki and E. D'Hoker, {\it Mod.\ Phys.\ Lett.\ }{\bf A7} (1992)
235.}
\REF\CF{Y. Tanii and S. Yamaguchi,
{\it Mod. Phys. Lett.} {\bf A6} (1991) 2271.}
\REF\DO{V.S. Dotsenko, {\it Mod. Phys. Lett.} {\bf A6} (1991) 3601.}
\REF\AG{L. Alvarez-Gaum\'e, J.L.F. Barb\'on and C. G\'omez,
{\it Nucl.\ Phys.\ }{\bf B368} (1992) 57.}
\Ref\STRING{A.M. Polyakov,
{\it Mod. Phys. Lett.} {\bf A6} (1991) 635;
D. Minic and Z. Yang, {\it Phys.\ Lett.\ }{\bf 274B} (1992) 27;
N. Sakai and Y. Tanii, {\it Phys. Lett.} {\bf 276B} (1992) 41;
to appear in {\it Prog.\ Theor.\ Phys.\ Suppl.\ }}
\REF\DFKU{P. Di Francesco and D. Kutasov,
{\it Phys.\ Lett.\ }{\bf 261B} (1991) 385;
{\it Nucl.\ Phys.\ }{\bf B375} (1992) 119.}
\Ref\OPEN{M. Bershadsky and D. Kutasov,
{\it Phys.\ Lett.\ }{\bf 274B} (1992) 331;
Y. Tanii and S. Yamaguchi,
{\it Mod.\ Phys.\ Lett.\ }{\bf A7} (1992) 521;
Saitama preprint STUPP--92--128 (March 1992).}
\REF\BPZ{A.A. Belavin, A.M. Polyakov and A.B. Zamolodchikov,
{\it Nucl.\ Phys.\ }{\bf B241} (1984) 333.}
\REF\DF{V.S. Dotsenko and V.A. Fateev,
{\it Nucl.\ Phys.\ }{\bf B240} (1984) 312; {\bf B251} (1985) 691;
V.S. Dotsenko, {\it Advanced Studies in Pure Mathematics} {\bf 16}
(1988) 123.}
\REF\GHMA{D. Ghoshal and S. Mahapatra,
Tata preprint TIFR--TH--91--58 (December, 1991).}
\REF\GJJ{S. Govindarajan, T. Jayaraman and V. John,
Madras preprint, IMSc--92/35 (August, 1992).}

%
%
\unnumberedchapters
\Pubnum={STUPP--92--132}
\titlepage
\title{\bf Correlation Functions of $(2k\!-\!1,2)$ Minimal Matter
           Coupled to 2D Quantum Gravity}
\vskip 5mm
\author{ Shun-ichi Yamaguchi }
\vskip 10mm
\address{ Physics Department, Faculty of Science \break
          Saitama University \break
          Urawa, Saitama 338, Japan }
\vskip 25mm
\abstract
We compute $N$-point correlation functions of non-unitary
$(2k\!-\!1,2)$ minimal matter coupled to 2D quantum gravity on a
sphere using the continuum Liouville field approach.
A gravitational dressing of the matter primary field with the
minimum conformal weight is used as the cosmological operator.
Our results are in agreement with the correlation functions of the
one-matrix model at the $k$-th critical point.
\endpage
%
%
Recent remarkable progress of 2D quantum gravity has been made by
the matrix model approach.\refmark{\BKDSGM,\GM}\ Matrix models
provided us important and interesting understanding of
non-perturbative aspects of 2D quantum gravity.
Another approach called the Liouville
approach\refmark{\KPZ,\DDK}\
uses a continuum field theory and also has been extensively
studied.\refmark{\SEPO-\OPEN}\
{}From the view point of the Liouville approach, it is important to
compute correlation functions in order to study the non-trivial
Liouville dynamics and in order to understand the precise
connection with the matrix models.
By using a free field approach and an analytic continuation
procedure, $N$-point correlation functions for general $N$
without the screening
charges in $c \le 1$ conformal matters coupled to 2D gravity
have been computed and analyzed in Refs.\ \STRING$-$\OPEN.\
They can be interpreted as scattering amplitudes of critical string
theories in 2D target space with non-trivial background fields.
In the case of $c < 1$ minimal matters\refmark{\BPZ,\DF}\ which
require the screening charges,
three-point functions have been computed.\refmark{\KITA,\DFKU}\
The results in unitary minimal matters were found to be consistent
with the matrix model results.
\par
In the Liouville approach, correlation functions of non-unitary
minimal matters are not sufficiently understood so far.
By comparing the gravitational dimensions\refmark{\KPZ,\DDK}\ of
physical operators in the Liouville approach and the matrix model
approach, one finds that a gravitational dressing of the matter
primary field with the minimum conformal weight should be used as
the cosmological operator.\refmark{\SEPO,\AG}\
For unitary matters the cosmological operator is the ordinary one
depending only on the Liouville field since the matter minimum
weight primary field is the identity operator.
On the other hand, for non-unitary minimal matters
one must use a modified cosmological operator which depends on
both of the matter and the Liouville fields.
\par
In Ref.\ \GHMA\ three-point functions of non-unitary $(2k\!-\!1,2)$
minimal matter\refmark\BPZ\ were computed.
However, two kinds of gravitational dressings of the
matter identity operator were used as cosmological operators as
in Ref.\ \DO. It is not clear to us whether such a method is
appropriate to non-unitary matters.
\par
The purpose of this letter is to compute $N$-point functions
of the non-unitary $(2k\!-\!1,2)$ minimal matter coupled to 2D
gravity on a sphere. They are obtained in the Liouville approach
using the action with the modified cosmological operator.
We use a similar technique as that used in Ref.\ \DFKU\
to compute $N$-point functions without screening charges.
Our results are in agreement with the correlation functions of the
one-matrix models at the $k$-th critical point,\refmark\GM\ which
are believed to represent the $(2k\!-\!1,2)$ minimal matter coupled
to 2D gravity.
\par
Recently, we received a preprint by Govindarajan
{\it et al.},\refmark\GJJ\ in which $N$-point functions of
the $(2k\!-\!1,2)$ matter were obtained using different method from
ours. They used algebraic properties of the BRST cohomology of
the theory. Our results are also consistent with theirs.
\par
We now consider the non-unitary $(2k\!-\!1,2)$ minimal
matter\refmark\BPZ\ coupled to 2D quantum gravity.
The matter conformal field theory has the central charge
$$
c = 1 - 12 \alpha_0^2\,,
\eqn\matterc$$
with
$$
\alpha_0 = -{1 \over 2} \alpha_+ + {1 \over 2}\alpha_-\,, \quad
\alpha_+ = -{2 \over \sqrt{2k-1}}\,, \quad
\alpha_- = -\sqrt{2k-1}\,.
\eqn\alp$$
After conformal gauge fixing, the system
is described by the matter field $X$ and the Liouville field
$\phi$ with the action \refmark{\DDK,\DF}
$$\eqalign{
S[\hat g,\phi,X] &= {1 \over 8\pi} \int d^2z \sqrt{\hat g}
\left(\hat g^{\alpha \beta} \partial_\alpha \phi \partial_\beta \phi
- Q \hat R \phi \right) + {\mu \over \pi}O_{\rm m} \cr
&\,\quad + {1 \over 8\pi} \int d^2z \sqrt{\hat g}
\left( \hat g^{\alpha \beta} \partial_\alpha X \partial_\beta X
+ 2i\alpha_0 \hat R X \right) + O_+ + O_- \,,
}\eqn\action$$
where $\hat R$ is the scalar curvature on a sphere
with a fixed reference metric $\hat g_{\alpha \beta}$ and
$\mu$ is the cosmological constant.
The operators $\,O_{\pm} = \int d^2z \sqrt{\hat g}
\e^{\mp i \alpha_\pm X}\,$
are the matter screening charges and
the operator $O_{\rm m}$ is the cosmological operator,
which will be defined later.
The parameter $Q$ is given by \refmark{\DDK}
$$
Q = - \alpha_+ - \alpha_- \,.
\eqn\Q$$
\par
For each matter primary field $\Psi_h = \e^{ipX}$,
which has a conformal weight
$h = {1 \over 2}p^2 -\alpha_0 p$,
there exists a physical operator\refmark{\DDK,\SEPO}
$$\eqalign{
& O^{(\pm)}_p
= \int d^2z \sqrt{\hat g} \e^{\beta^{(\pm)} \phi \,+\, ipX}\,, \cr
& \quad \beta^{(\pm)}(p) = -{Q \over 2} +|p-\alpha_0|
= \left\{ \matrix{ p + \alpha_+ \quad \cdots \quad
{\rm for} \ \ p \geq \alpha_0 \hfill \cr
-p + \alpha_- \quad \cdots \quad {\rm for} \ \ p <
\alpha_0 \hfill \cr} \right.\,.
}\eqn\phyop$$
For the $(2k\!-\!1,2)$ matter
the momentum $p$ of matter primary fields
takes discrete values\refmark{\BPZ,\DF}\ parametrized by an
integer $t$
$$
p=\left\{\matrix{
p_t = {1 \over 2}(t-1)\alpha_+ \ \geq \alpha_0 \hfill \cr
\overline{p}_t = \alpha_- - {1 \over 2}(t + 1)\alpha_+ \ <
\alpha_0 \hfill
}\right. \qquad (1 \leq t \leq k-1)\,.
\eqn\minimomenta$$
Operators $O_{p_t}^{(+)}$ and $O_{\overline{p}_t}^{(-)}$ have
the same matter conformal weight and represent
the same operator of the minimal matter.
The corresponding Liouville momenta are
$\beta^{(+)}(p_t)=\beta^{(-)}(\overline{p}_t)={1 \over 2}(t+1)
\alpha_+\,$. The cosmological operator $O_{\rm m}$
in Eq.\ \action\ is\refmark{\SEPO,\AG}
$$
O_{\rm m} = \int d^2 z \sqrt{\hat g}
\e^{\beta_{\rm m} \phi} \Psi_{h_{\rm m}}\,,
\eqn\mcosmo$$
where
$\Psi_{h_{\rm m}}
\,(= \!\e^{i p_{\rm m} X}\ {\rm or}\
\e^{i \,\overline{p}_{\rm m} X})$ is a matter primary field with
the minimum conformal weight $h_{\rm m}$.
It has a momentum $p_{\rm m}=p_{t=k-1}$ or
$\overline{p}_{\rm m}=\overline{p}_{t=k-1}$ in Eq.\ \minimomenta\
and $\beta_{\rm m}\,( = \beta_{\rm m}^{(+)} = \beta_{\rm m}^{(-)})$
in Eq.\ \phyop.\
The cosmological operator $O_{\rm m}$ is
one of the physical operators in the theory.
\par
We consider the $N$-point functions
of the operators \phyop\ with momenta \minimomenta\ on a sphere.
As in Refs.\ \KITA,\ \DFKU,\
we first integrate the Liouville  zero mode
$\phi_0$ $(\phi = \phi_0 + \tilde \phi)$
and define $s_{\rm m}$ by
$$
\sum_{i=1}^N \beta_i^{(+)}
+ s_{\rm m} \beta_{\rm m} = -Q \,.
\eqn\lzmcon$$
Eq.\ \lzmcon\ can be regarded as the conservation of the Liouville
momentum. Next, we shall integrate the matter zero mode
$X_0$ $(X = X_0 + \tilde X)$. Note that a factor
$\left(
\int d^2z \sqrt{\hat g} \e^{\beta_{\rm m} \tilde \phi}
\Psi_{h_{\rm m}} \right)^{s_{\rm m}}$
resulting from the $\phi_0$ integration
contributes to the matter zero mode integration.
In order to simplify the matter momentum conservation and
to symmetrize our final results under permutations of
$(t_1,t_2,\cdots,t_N)$, without loosing generality
we can use
$s_{\rm m}\!-\!1$ $\e^{i p_{\rm m} X}$'s
and one $\e^{i \,\overline{p}_{\rm m} X}$
for $s_{\rm m}$ $\Psi_{h_{\rm m}}$'s.
These two types of operators represent the same operator
$\Psi_{h_{\rm m}}\,$.
Then the matter momentum conservation can be satisfied
by inserting $n$ screening charges $O_+$ and no $O_-$
$$
\left[ \, \sum_{i=1}^N p_i + (s_{\rm m} \!-\! 1)p_{\rm m}
+ \overline{p}_{\rm m} \,\right]
- n \alpha_+ = -\alpha_+ + \alpha_- \,.
\eqn\mzmcon$$
Thus we obtain the result after the zero mode $X_0,\phi_0$
integrations
$$\eqalign{
&\VEV{O_1^{(+)}\cdots O_N^{(+)}} \cr
&= 2 \pi \delta \left(
\sum_{i=1}^N p_i + (s_{\rm m} \!-\! 1)p_{\rm m}
+ \overline{p}_{\rm m} +(1-n)\alpha_+ - \alpha_- \right)
{\Gamma(-s_{\rm m}) \over |\beta_{\rm m}|}
\left( {\mu \over \pi} \right)^{s_{\rm m}} \cr
&\quad\ \times
\VEV{\tilde O_1^{(+)}(0)\,\tilde O_2^{(+)}(\infty)\,\tilde
O_3^{(+)}(1)\,
\tilde O_4^{(+)} \,\cdots \,\tilde O_N^{(+)}\,
\left[ \left( \tilde O_{{\rm m}}^{(+)} \right)^{s_{\rm m}-1}
\tilde O_{{\rm m}}^{(-)} \right]
{1 \over n!} \left(\tilde O_+ \right)^n },
}\eqn\nun$$
where we have used shorthand notations
$\,\tilde O_i(z) = \e^{\beta_i \tilde \phi \,+\, i p_i \tilde X}$,
$\tilde O_i = \int d^2z \sqrt{\hat g} \,\tilde O_i(z)$ etc.\
and fixed $SL(2,{\bf C})$ gauge symmetry by setting
$z_1 = 0, z_2 = \infty, z_3 = 1$.
The non-zero mode $\tilde \phi, \tilde X$ expectation value
on the right hand side of Eq.\ \nun\ is
defined by using the free field action.
\par
To evaluate the non-zero mode
expectation value in Eq.\ \nun,
let us consider more general cases
$$
\VEV{\tilde O_1^{(+)}(0)\,\tilde O_2^{(+)}(\infty)\,\tilde
O_3^{(+)}(1)
\,\tilde O_4^{(+)} \,\cdots \,\tilde O_{M-1}^{(+)} \tilde O_M^{(-)}
\left(\tilde O_c \right)^s {1 \over n!}
\left(\tilde O_+ \right)^n } \,,
\eqn\cnpt$$
where $\,O_c = \int d^2z \sqrt{\hat g} \e^{\alpha_+ \phi}\,$
is a physical operator $O^{(+)}_{p=0}$ in Eq.\ \phyop.
The matter momentum conservation and the definition of $s$ are
$$
\sum_{i=1}^M p_i -n\alpha_+ = -\alpha_+ + \alpha_-\,, \qquad
\sum_{i=1}^M \beta_i + s \alpha_+ = -Q \,.
\eqn\ccon$$
The expectation value in Eq.\ \nun\ corresponds to $s\!=\!0$,
$M\!=\!N\!+\!s_{\rm m}$ and $p_i\!=\!p_{\rm m}$
$(i\!=\!N\!+\!1,\cdots,N\!+\!s_{\rm m}\!-\!1)$,
$p_{i=N+s_{\rm m}}\!=\!\overline{p}_{\rm m}\,$.
Eq.\ \cnpt\ for $M\!=\!3$ was
evaluated in Ref.\ \DFKU\ (see also Ref.\ \KITA)
using the techniques of analytic continuations
in the momenta and the matter central charge.
The non-zero mode integration is performed by tuning
the momenta $p_1,p_2,p_3$
and the matter central charge $c$
so that $s$ and $n$ are non-negative integers.
The expectation value with desired values of
the momenta and the central charge is
then obtained by an analytic continuation.
In this way the three-point function was obtained as\refmark{\DFKU}\
$$\eqalign{
&\VEV{\tilde O_1^{(+)}(0)\,\tilde O_2^{(+)}(\infty)\,\tilde
O_3^{(-)}(1) \left(\tilde O_c \right)^s
{1 \over n!} \left(\tilde O_+ \right)^n } \cr
&\qquad\quad\qquad = \pi^{s+n}
{\Gamma(s+1) \over \bigl[\,\Gamma(s+n+1)\,\bigr]^2}
\,[\Delta(-\rho)]^s \,[\Delta(\rho)]^n
\prod_{i=1}^2 \Delta(m_i) \,,
}\eqn\three$$
where
$$\rho = -{1 \over 2}\alpha_+^2\,, \quad
m_i = m(p_i) = {1 \over 2} \beta_i^2 - {1 \over 2} p_i^2\,, \quad
\Delta(x) = {\Gamma(x) \over \Gamma(1-x)}\,.
\eqn\m$$
In Ref.\ \DFKU\ $M$-point functions \cnpt\
for general $M$
without screening charge $n\!=\!0$ were also obtained.
It was pointed out that
the matter screening charges should not be treated on the equal
footing as physical operators \phyop.\
For example, using the analytic continuation procedure,
the expectation value \cnpt\ for $M\!=\!3$
with $n$ screening charges
can not be obtained
from that of $M\!=\!3\!+\!n$ with no screening charge.
\par
We shall now obtain the four-point function.
We closely follow the procedure of Ref.\ \DFKU\ to derive
four-point functions from three-point functions
when there is no screening charge.
{}From the kinematical constraints \ccon\ $m_i$ satisfy
$$
\sum_{i=1}^3 m_i = (s-n) \rho +1 \,, \qquad m_4=-s-n-1\,,
\eqn\zmconfour$$
and we can take $m_1$ and $m_3$ as independent variables.
Eq.\ \cnpt\ for $M\!=\!4$ and $s,n \in {\bf Z_+}$ becomes
$$\eqalign{
&\VEV{\tilde O_1^{(+)}(0)\,\tilde O_2^{(+)}(\infty)\,\tilde
O_3^{(+)}(1) \,\tilde O_4^{(-)}
\left(\tilde O_c \right)^s
{1 \over n!} \left(\tilde O_+ \right)^n } \cr
& \qquad = {1 \over n!} \int d^4 z_4 \,|z_4|^{2a_1}
\,|1-z_4|^{2a_3} \cr
& \qquad\quad \ \times \int \prod_{i=1}^s \left[\,d^2 w_i
\,|w_i|^{2b_1^+}
\,|1-w_i|^{2b_3^+}
\,|z_4-w_i|^{2c^-} \right]
\prod_{1 \leq i < j \leq s} |w_i-w_j|^{4\rho} \cr
& \qquad\quad \ \times \int \prod_{i=1}^n \left[\,d^2 u_i
\,|u_i|^{2b_1^-}
\,|1-u_i|^{2b_3^-}
\,|z_4-u_i|^{2c^+} \right]
\prod_{1 \leq i < j \leq n} |u_i-u_j|^{-4\rho} \cr
& \qquad \equiv F(m_1,m_3;\rho,s,n) \,,
}\eqn\F$$
where
$$a_i = (s+n+1)m_i -1\,, \quad
b_i^{\pm} = -m_i \pm \rho\,, \quad
c^{\pm} = \pm(s+n+1)\rho -1\,.
\eqn\abc$$
\par
We first consider $F$ as a function of $m_1$ and discuss its
analyticity. From the integral representations \F,\
one can see that poles in $m_1$
arise when some of the integration variables $z_4,w_i,u_i$
approach to the origin or infinity.
Let us first examine the poles arising from
integration regions near the origin.
The positions of these poles are independent of $m_3$.
Therefore we can find them by setting $p_3=0$,
which reduces \F\ to the three point function \three\
with $s \rightarrow s+1$ for the operators
$\tilde O_1$, $\tilde O_2$ and $\tilde O_4$.
Thus, from Eq.\ \three\ we find that the poles arise only when
$\tilde O_4$ approaches to $\tilde O_1$ $(z_4 \rightarrow 0)$
and their positions are $m_1=0,-1,-2,\cdots$.
Next we shall examine the poles arising from
integration regions near infinity.
They can be naturally understood in terms of
$\,m_2 = -m_1-m_3+(s-n)\rho+1\,$.
By changing the integration variables
$z\,(=\!z_4,w_i,u_i) \rightarrow 1/z$,
we find that these poles arise when $\tilde O_4$ approaches to
$\tilde O_2$ and their positions are $m_2=0,-1,-2\cdots$.
By similar arguments we find that poles in $m_3$
arise at $m_3 = 0,-1,-2\cdots$.
There is no other singularity in Eq.\ \F.\
\par
Let us consider the quantity
$$
f(m_1,m_3;\rho,s,n)
={F(m_1,m_3;\rho,s,n) \over \Delta(m_1)\,\Delta(m_2)\,
\Delta(m_3)}\,.
\eqn\sf$$
We can show that $f(m_1,m_3;\rho,s,n)$ is independent of
$m_1$ and $m_3$ as follows.\refmark\DFKU\ Changing the integration
variables $\,z'\,(=z'_4,w'_i,u'_i) = m_1 \ln z\,$ in
Eq.\ \F,\ we find that $F \sim |m_1|^{-2m_3 + 2(s-n)\rho}$
for large $|m_1|$ and therefore $f$ is a constant for
$|m_1| \rightarrow \infty$.
The singularities of $F$ in $m_1$ are canceled by those of
$\Delta (m_i)$.
Furthermore, it can be shown in the same way as in Ref.\ \DFKU\
that $F$ vanishes at $m_1 = 1, 2, 3, \cdots$,
which are zeros of the denominator in Eq.\ \sf.
Thus we find that $f$ is an analytic function on the whole complex
$m_1$-plane including a point at infinity.
Therefore it must be a constant independent of $m_1$.
Similarly $f$ is shown to be independent of $m_3$.
The explicit form of $f(\rho,s,n)$ is obtained as follows.
If we put one of the momenta to zero in Eq.\ \F,
it is reduced to a three-point function with $s\!+\!1$ $\tilde
O_c$'s. By comparing with Eq.\ \three,\ we obtain
$$
f(\rho,s,n) = \pi^{s+n+1}
{\Gamma(s+2) \over \bigl[\,\Gamma(s+n+2)\,\bigr]^2}
\,[\Delta(-\rho)]^s \,[\Delta(\rho)]^n \,.
\eqn\fresult$$
{}From Eqs.\ \sf\ and \fresult\ we obtain
$$
\VEV{\tilde O_1^{(+)}(0)\,\tilde O_2^{(+)}(\infty)\,
\tilde O_3^{(+)}(1)
\,\tilde O_4^{(-)}
\left(\tilde O_c \right)^s
{1 \over n!} \left(\tilde O_+ \right)^n }
= f(\rho,s,n)\,\prod_{i=1}^3 \Delta(m_i)\,.
\eqn\four$$
\par
We can easily generalize above arguments
to $M$-point $(M \!\geq\! 5)$ functions.
The variables $m_i$ satisfy
$$
\sum_{i=1}^{M-1} m_i = (s-n) \rho +1\,,\qquad m_M=-s-n-M+3\,.
\eqn\ncon$$
The analyticity in $m_i$ and comparison with lower point functions
give the result\foot
{This result is also useful for unitary minimal matters
coupled to 2D gravity, in which the cosmological operator
is $O_c$.}
$$\eqalign{
&\VEV{\tilde O_1^{(+)}(0)\,\tilde O_2^{(+)}(\infty)\,
\tilde O_3^{(+)}(1)
\,\tilde O_4^{(+)} \,\cdots \,\tilde O_{M-1}^{(+)} \tilde O_M^{(-)}
\left(\tilde O_c \right)^s {1 \over n!}
\left(\tilde O_+ \right)^n } \cr
&\qquad = \pi^{s+n+M-3}
{\Gamma(s+M-2) \over \bigl[\,\Gamma(s+n+M-2)\,\bigr]^2 }\,
\bigl[\Delta(-\rho) \bigr]^s \bigl[\Delta(\rho) \bigr]^n
\prod_{i=1}^{M-1} \Delta(m_i) \,.
}\eqn\npt$$
\par
Let us now obtain
the non-zero mode expectation values in Eq.\ \nun\ using the above
results. By setting $s\!=\!0$, $M\!=\!N\!+\!s_{\rm m}$ and
$p_i\!=\!p_{\rm m}$ $(i\!=\!N\!+\!1,\cdots,N\!+\!s_{\rm m}\!-\!1)$,
$p_{i=N+s_{\rm m}}\!=\!\overline{p}_{\rm m}$
in Eq.\ \npt,\ we obtain
$$\eqalign{
&\VEV{\tilde O_1^{(+)}(0)\,\tilde O_2^{(+)}(\infty)\,
\tilde O_3^{(+)}(1)
\,\tilde O_4^{(+)}\,\cdots \,\tilde O_N^{(+)}
\left[ \left( \tilde O_{{\rm m}}^{(+)} \right)^{s_{\rm m}-1}
\tilde O_{\rm m}^{(-)} \right]
{1 \over n!} \left(\tilde O_+ \right)^n } \cr
&\qquad = \pi^{s_{\rm m}+n+N-3}{\Gamma(s_{\rm m}+N-2)
\over \bigl[\,\Gamma(s_{\rm m} +n +N -2)\,\bigr]^2} \,
[\Delta(\rho)]^n
[\Delta(m_{\rm m})]^{s_{\rm m}-1}
\,\prod_{i=1}^N \Delta(m_i) \,,
}\eqn\nznu$$
where
$m_{\rm m} = {1 \over 2} \beta_{\rm m}^2 - {1 \over 2}
p_{\rm m}^2\,$.
The result \nznu\ obtained for non-negative integers $s_{\rm m}$
is now analytically continued to a desired value of $s_{\rm m}$.
Therefore, by Eq.\ \nznu\ we find that
the $N$-point function \nun\ becomes
$$\eqalign{
\VEV{O_1^{(+)}\cdots O_N^{(+)}}
& = 2 \pi \delta \left(
\sum_{i=1}^N p_i + (s_{\rm m} \!-\! 1)p_{\rm m}
+ \overline{p}_{\rm m} +(1-n)\alpha_+ - \alpha_- \right) \cr
&\quad \ \times  {\mu^{s_{\rm m}} \over |\beta_{\rm m}|} \,
{ \Gamma(s_{\rm m}+N-2) \over \Gamma(s_{\rm m}+1) } \,
(-\pi)^{n+N-3} [\Delta(\rho)]^n \cr
&\quad \ \times [\Delta(m_{\rm m})]^{s_{\rm m} -1} \,
\Delta(-s_{\rm m} -n-N+3)
\prod_{i=1}^N \Delta(m_i) \,.
}\eqn\nunreslt$$
{}From Eqs.\ \lzmcon\ and \mzmcon\ we obtain
$$
s_{\rm m} = {1 \over k} \left( 2k+1-N-\sum_{i=1}^{N}t_i \right)
\,,\qquad
n= -s_{\rm m}-N+{5 \over 2}\,,
\eqn\smn$$
where $t_i$ is parameters for momenta $p_i$ in Eq.\ \minimomenta.\
The parameters defined in Eq.\ \m\ take the values
$$
\rho = -{2 \over 2k-1}\,,\quad
m_i = {2 t_i \over 2k-1} \,,\quad
m_{\rm m} = {2k-2 \over 2k-1}\,.
\eqn\frpara$$
Substituting Eqs.\ \smn\ and \frpara\ into Eq.\ \nunreslt\
we obtain the final result of the $N$-point function.
\par
Let us compare this result with that of the matrix model approach.
The normalizations of the partition function and the physical
operators may be different in two approaches. We rescale the
partition function and renormalize the operators and the
cosmological constant by finite real factors as
$$\eqalign{
Z\ \ &\rightarrow\ \ -\bigl[\pi (1-2k) \bigr]^{-{1 \over 2}}
\left[ \Delta\!\left( {2k+1 \over 2k-1} \right)
\right]^{{5 \over 2}}
\Delta\!\left( {2k-2 \over 2k-1} \right)
\,Z \,,\cr
O_{p_{t_i}}^{(+)}\,&=\,\Delta\!\left( {2k+1 \over 2k-1} \right)
\Delta\!\left({2 t_i \over 2k-1}\right) \,O_{l_i}
\qquad (\,l_i \equiv k-1-t_i \,)\,, \cr
\mu\,&=\,-\pi \,\Delta\!\left(-{2 \over 2k-1} \right)
\Delta\!\left({1 \over 2k-1} \right) \,\mu_r\,,
}\eqn\rescale$$
and obtain
$$
\VEV{O_{l_1}\cdots O_{l_N}}\,
= \,-{1 \over k} \,
{\Gamma \!\left[ \left( \sum_{i=1}^N l_i +1 \right)/k \right] \over
\Gamma \!\left[ \left( \sum_{i=1}^N l_i +1 -(N-3)k \right)/k
\right]}
\,{\mu_r}^{\left[\sum_{i=1}^N l_i +1 -(N-2)k \right]/k }\,.
\eqn\fr$$
The integer parameters $l_i$
take values $0,1,\cdots,k\!-\!2\,$.
This result is in agreement with
the correlation functions of the one-matrix model
at the $k$-th critical point.\refmark\GM\foot
{There are misprints in the formula for the correlation functions
(Eq.\ (4.19)) given in the first paper of Ref.\ \GM.}
\bigskip
\par
\ack
The author would like to thank Y. Tanii for many valuable
discussions and very careful reading of this manuscript.
He is also grateful to members of the theory group
at Saitama University for useful comments and encouragements.
\bigskip
\refout
\end